\documentclass{article}

\PassOptionsToPackage{numbers}{natbib}

\usepackage[final]{score_workshop}

\usepackage[utf8]{inputenc} 
\usepackage[T1]{fontenc}    
\usepackage{hyperref}       
\hypersetup{
    colorlinks,
    linkcolor={red!80!black},
    citecolor={blue!80!black},
    urlcolor={blue!60!black}
}
\usepackage{url}            
\usepackage{booktabs}       
\usepackage{amsfonts}       
\usepackage{nicefrac}       
\usepackage{microtype}      
\usepackage{xcolor}         

\usepackage{amsmath,amsfonts,bm,amssymb,mathtools,amsthm}
\usepackage{color,xcolor,xspace}
\usepackage{booktabs}
\usepackage{thm-restate}
\usepackage{bbding}
\usepackage{pifont}
\usepackage{wasysym}
\usepackage[most]{tcolorbox}

\newcommand{\bb}[1]{{\mathbb{#1}}}
\newcommand{\norm}[1]{{\lVert {#1} \rVert}}

\def\eqref#1{Eq.(\ref{#1})}

\def\1{\bm{1}}

\def\rvx{{\mathbf{x}}}
\def\rvy{{\mathbf{y}}}
\def\rvz{{\mathbf{z}}}

\def\mH{{\bm{H}}}
\def\mI{{\bm{I}}}

\DeclareMathAlphabet{\mathsfit}{\encodingdefault}{\sfdefault}{m}{sl}
\SetMathAlphabet{\mathsfit}{bold}{\encodingdefault}{\sfdefault}{bx}{n}

\def\gN{{\mathcal{N}}}

\def\sR{{\mathbb{R}}}

\definecolor{first}{rgb}{0.1, 0.42, 0.1}
\definecolor{second}{rgb}{0.0, 0.35, 0.56}

\title{JPEG Artifact Correction using \\ Denoising Diffusion Restoration Models}

\author{%
  Bahjat Kawar\thanks{Equal contribution.} \\
  Technion, Haifa, Israel\\
  \texttt{bahjat.kawar@cs.technion.ac.il} \\
  \And
  Jiaming Song$^*$ \\
  NVIDIA\\
  \texttt{jiamings@nvidia.com} \\
  \And
  Stefano Ermon \\
  Stanford University, CA\\
  \texttt{ermon@cs.stanford.edu} \\
  \And
  Michael Elad \\
  Technion, Haifa, Israel\\
  \texttt{elad@cs.technion.ac.il} \\
}

\begin{document}

\maketitle

\begin{abstract}
Diffusion models can be used as learned priors for solving various inverse problems. However, most existing approaches are restricted to linear inverse problems, limiting their applicability to more general cases. In this paper, we build upon Denoising Diffusion Restoration Models (DDRM) and propose a method for solving some non-linear inverse problems. We leverage the pseudo-inverse operator used in DDRM and generalize this concept for other measurement operators, which allows us to use pre-trained unconditional diffusion models for applications such as JPEG artifact correction. We empirically demonstrate the effectiveness of our approach across various quality factors, attaining performance levels that are on par with state-of-the-art methods trained specifically for the JPEG restoration task.
Our code is available at \url{https://github.com/bahjat-kawar/ddrm-jpeg}.
\end{abstract}

\begin{figure}[h]
    \centering
    \vspace{-0.2cm}
    \includegraphics[width=0.76\textwidth]{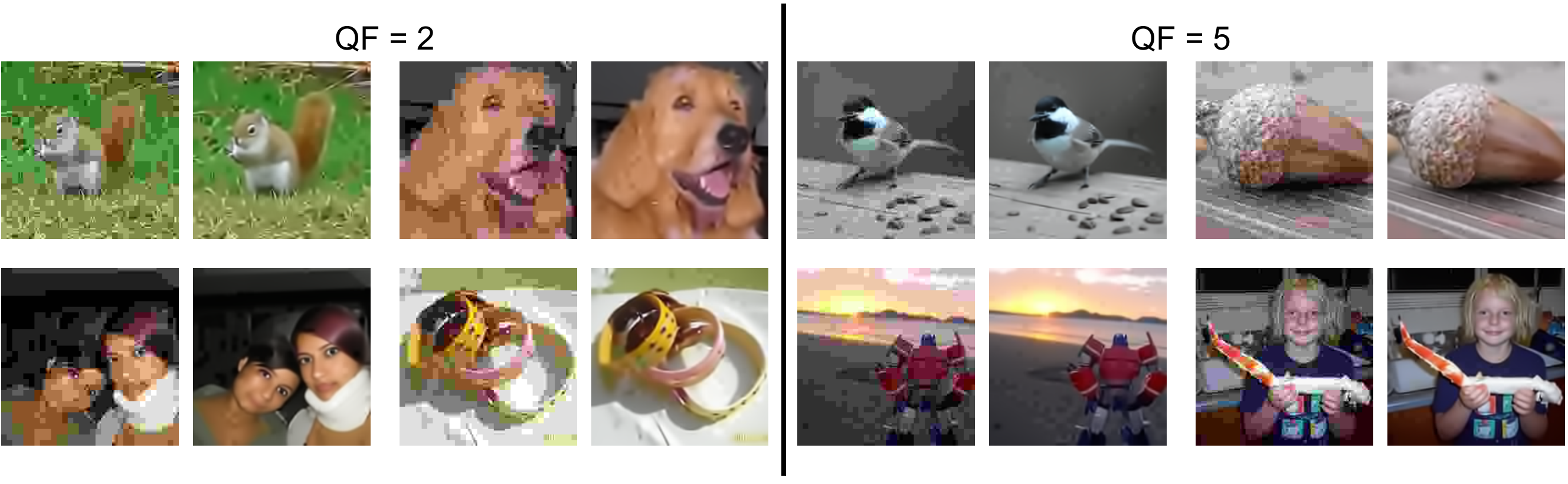}
    \vspace{-0.2cm}
    \caption{Pairs of JPEG images and restorations using our method. Best viewed zoomed in.}
    \label{fig:headline}
\end{figure}

\section{Introduction}
Many image processing problems are instances of inverse problems~\cite{ledig2016photo,kupyn2019deblurgan,larsen2016autoencoding}. In real-world applications, one would often need to face multiple different degradation models~\cite{song2021medical,jalal2021robust,ddrm}, where training problem-specific models on each case could be expensive~\cite{palette}. Therefore, it is valuable to develop methods that apply problem-agnostic models, which would adapt to different degradation models at inference time without retraining. Existing approaches, while achieving decent performance on a variety of tasks, are generally limited to linear inverse problems~\cite{ilvr,song2020score,snips,ddrm}, leaving out certain important non-linear inverse problems such as JPEG artifact correction.
Since JPEG is a lossy image compression format~\cite{wallace1991jpeg}, JPEG images exhibit loss of quality and undesired artifacts. Several methods have been developed for addressing this problem~\cite{xiong1997deblocking, bruckstein2003down, dong2015compression, liu2018multi, qgac, jiang2021towards}.

To address this issue, we introduce a method that performs JPEG artifact correction using Denoising Diffusion Restoration Models (DDRM)~\cite{ddrm}. Our core idea is to generalize the pseudo-inverse matrix that exists in the DDRM update rule for the noiseless observation case. This generalized notion of a ``pseudo-inverse'' includes JPEG as a special case, where the ``pseudo-inverse'' for JPEG encoding is simply JPEG decoding. The resulting algorithm resembles the original update for DDRM, replacing the linear operator and its pseudo-inverse with JPEG encoding and decoding, respectively. 

We apply our algorithm for JPEG restoration with various quality factors (QF), where the quantization matrices are embedded in the JPEG files and naturally known at inference time.
In common image quality metrics such as PSNR, SSIM~\cite{ssim}, and LPIPS~\cite{lpips}, our method compares favorably against a recent state-of-the-art GAN-based baseline~\cite{qgac} trained specifically for JPEG restoration. Our method achieves even more improvement on low QF that the baseline is not trained on, demonstrating the generalization advantages of methods that leverage unconditional diffusion models.

\begin{figure}
    \centering
    \begin{minipage}{0.64\textwidth}
        \centering
        \includegraphics[width=0.85\textwidth]{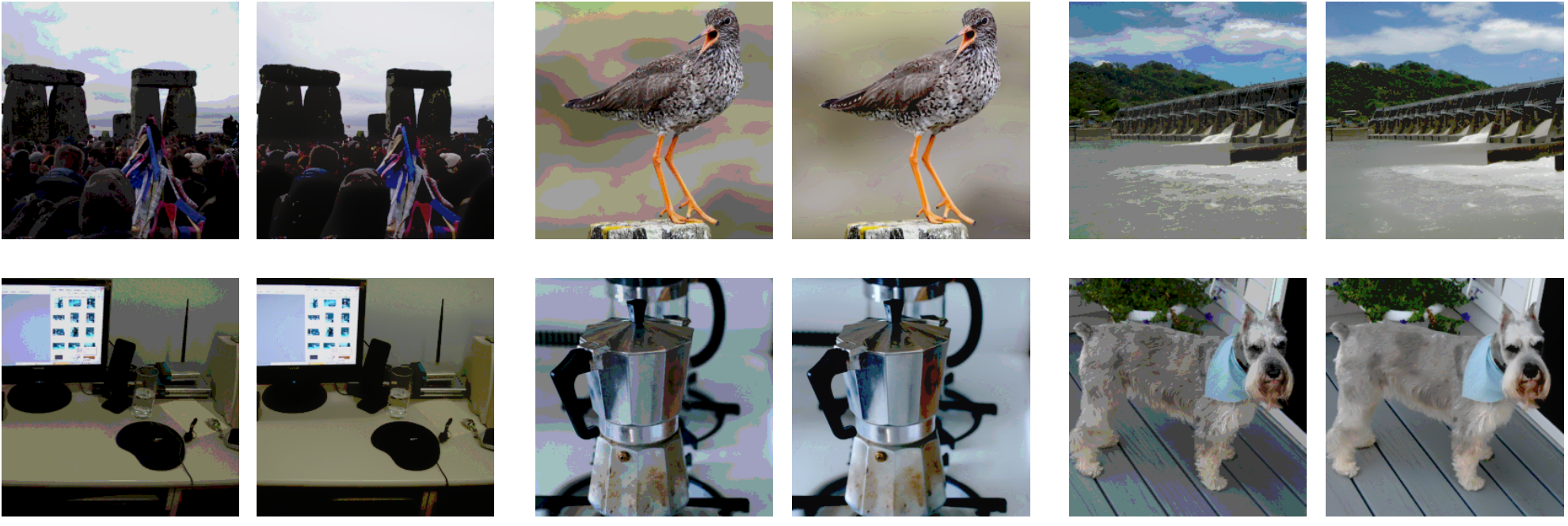}
        \caption{Pairs of quantized ($9$ bits per color) and restored images using our method. Best viewed zoomed in.}
        \label{fig:quant}
    \end{minipage}\hfill
    \begin{minipage}{0.33\textwidth}
        \centering
        \includegraphics[width=0.85\textwidth]{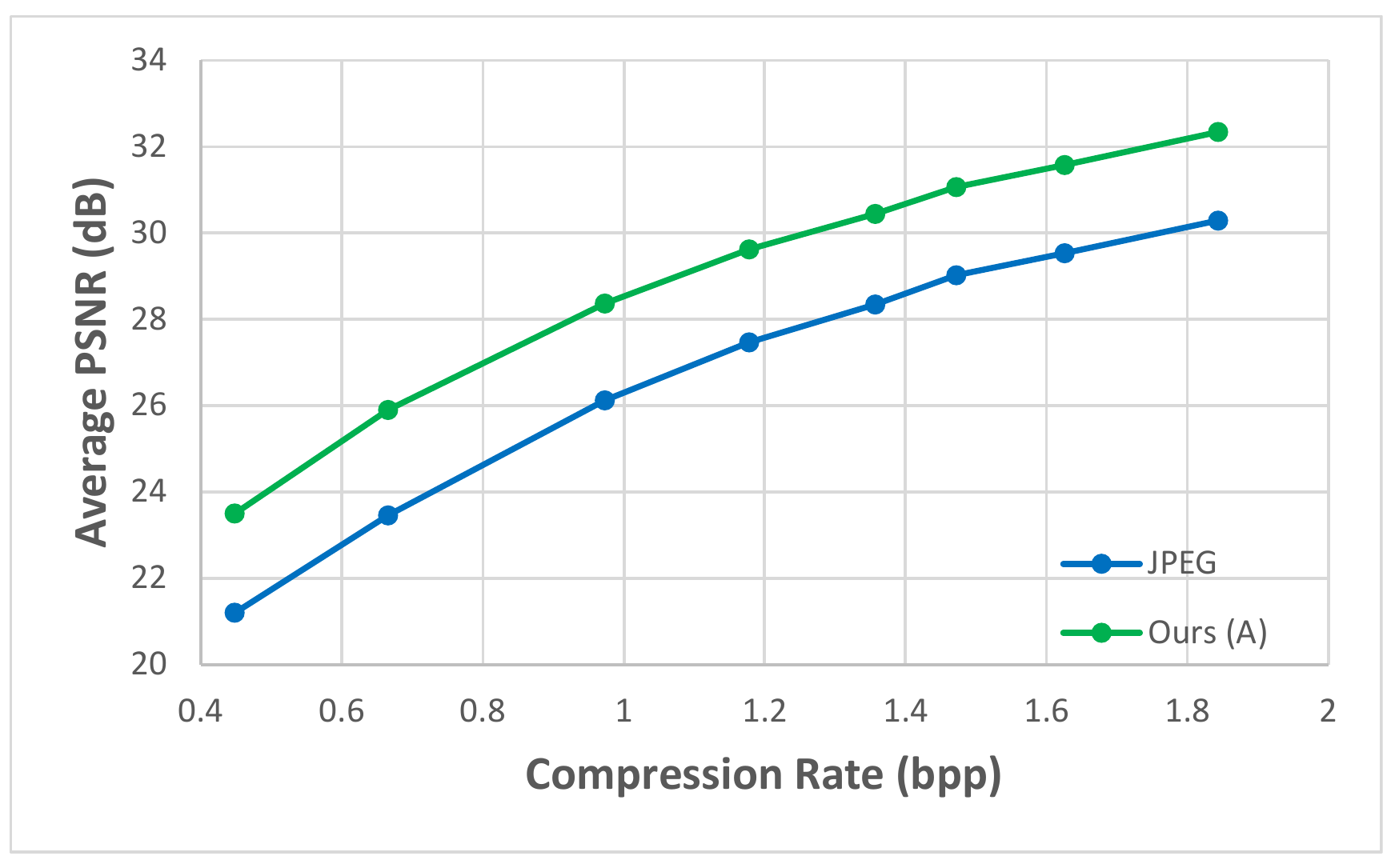}
        \caption{Rate-distortion curves for standard JPEG (blue) and our method (green).}
        \label{fig:rd}
    \end{minipage}
\end{figure}

\section{Background}

\paragraph{Diffusion Models.}
Diffusion models~\cite{sohl-dickstein2015deep, ddpm} are generative models with a Markov chain structure ${\rvx_T \to \rvx_{T-1} \to \ldots \to \rvx_{1} \to \rvx_{0}},$
where $\forall t: \rvx_t \in \sR^n$, which defines the following joint distribution:
$$p_\theta(\rvx_{0:T}) = p_\theta^{(T)}(\rvx_T) \prod_{t=0}^{T-1} p_\theta^{(t)}(\rvx_t | \rvx_{t+1}).$$
After $\rvx_{T}, \dots, \rvx_{0}$ are drawn, only $\rvx_{0}$ is kept as the final output of the generative model.
When training a diffusion model, a fixed, factorized variational inference distribution is assumed:
$$q(\rvx_{1:T} | \rvx_0) =  q^{(T)}(\rvx_T | \rvx_0) \prod_{t=0}^{T-1}q^{(t)}(\rvx_t | \rvx_{t+1}, \rvx_0),$$
which leads to an evidence lower bound (ELBO) on the maximum likelihood objective. One particular parametrization~\cite{song2020denoising} takes the following form:
\begin{align}
    q(\rvx_t | \rvx_0) = \gN(\sqrt{\alpha_t} \rvx_0, (1 - \alpha_t) \mI), \quad \forall t \in [1, T]
\end{align}
where $p_\theta^{(t)}$ can be trained via a denoising autoencoder~\cite{vincent2011connection} objective; in the ideal case, the denoiser, denoted as $f_\theta^{(t)}(\rvx_t)$ should map to the MMSE (Minimum Mean Squared Error) estimator $\bb{E}_{q(\rvx_0 | \rvx_t)}[\rvx_0]$, and constitutes a prediction over the ``clean'' $\rvx_0$.
Diffusion models have achieved unprecedented success in image generation~\cite{song2020denoising, vahdat2021score, guided_diffusion, kawar2022enhancing}, and they have also been deployed for a variety of tasks~\cite{kawar2021stochastic, de2021diffusion, nie2022DiffPure, blau2022threat, han2022card, amit2021segdiff, avrahami2022blended, shi2022conditional,sinha2021d2c}.

\paragraph{Linear Inverse Problems.}
A general linear inverse problem is posed as
\begin{align}
    \rvy = \mH \rvx + \rvz, \label{eq:inverse-problem-def}
\end{align}
where our aim is to recover the signal $\rvx \in \sR^{n}$ from measurements $\rvy \in \sR^{m}$. $\mH \in \sR^{m \times n}$ is a known degradation matrix, and $\rvz \sim \gN(0, \sigma_{\rvy}^2 \mI)$ is an additive white Gaussian noise with known variance.

Various works have applied diffusion models for inverse problem solving, mostly for the noiseless case. While it is possible to train a conditional diffusion model based on pairs of $\rvx$ and $\rvy$~\cite{palette,saharia2021image,whang2021deblurring}, such models may not generalize to other inverse problems. Therefore, it is often desirable to formulate inverse problem solvers from unconditional diffusion models~\cite{ilvr,song2020score,snips,ddrm}, where the knowledge about the inverse problem does not need to be known during training; compared with problem-specific conditional diffusion models, problem-agnostic techniques save significant computational resources. 

\paragraph{Denoising Diffusion Restoration Models (DDRM).}
In particular, DDRM~\cite{ddrm} is a general solver for linear inverse problems in both noisy and noiseless cases. For any linear inverse problem, the DDRM model is defined as 
$$p_\theta(\rvx_{0:T} | \rvy) = p_\theta^{(T)}(\rvx_T | \rvy) \prod_{t=0}^{T-1} p_\theta^{(t)}(\rvx_t | \rvx_{t+1}, \rvy),$$
where $\rvx_0$ is the final diffusion output. The high-level idea behind DDRM is to leverage the singular value decomposition of $\mH$ and transform both $\rvx$ and the possibly noisy $\rvy$ to a shared spectral space. In this space, DDRM performs denoising on dimensions when information from $\rvy$ is available (\textit{i.e.}, when singular values are non-zero) and performs imputation on ones where such information is not available (\textit{i.e.}, when singular values are zero), taking account of the measurement noise explicitly. 
\paragraph{JPEG.} JPEG~\cite{wallace1991jpeg} is a commonly used lossy compression method for images. At a high-level, JPEG first transforms an uncompressed image from the RGB color space to the YCbCr space, optionally applies chroma subsampling, splits the image into $8\times 8$ pixel blocks, performs a discrete cosine transform (DCT), and then performs quantization of the resulting values using a fixed quantization matrix. These values can then be compressed in a lossless fashion via Huffman trees.
The entire process can be reverted to define the JPEG decoding method, with loss of information happening in the chroma subsampling and the quantization steps.
Since its introduction in 1991, JPEG has become the most widely used image compression format in the world, with several billion JPEG images produced every day; thus, restoring high-quality images from JPEG-compressed ones has wide applications.
Several previous methods were developed for this purpose~\cite{xiong1997deblocking, bruckstein2003down, dong2015compression, liu2018multi, qgac, jiang2021towards}.

\begin{table}
  \caption{JPEG artifact correction results on ImageNet-1K for varying quality factors (QF).}
  \label{tab:imagenet}
  \centering
  \begin{tabular}{l ccc ccc ccc}
    \toprule
    &
    \multicolumn{3}{c}{\textbf{QF = }$\mathbf{5}$} &
    \multicolumn{3}{c}{\textbf{QF = }$\mathbf{10}$} &
    \multicolumn{3}{c}{\textbf{QF = }$\mathbf{30}$}
    \\
    \cmidrule(r){2-4}
    \cmidrule(r){5-7}
    \cmidrule(r){8-10}
    \textbf{Method}     &
    \textbf{PSNR} & \textbf{SSIM} & \textbf{LPIPS} &
    \textbf{PSNR} & \textbf{SSIM} & \textbf{LPIPS} &
    \textbf{PSNR} & \textbf{SSIM} &  \textbf{LPIPS}
    \\
    \midrule
    JPEG &
    $23.46$ & $0.62$ & $0.48$ &
    $26.12$ & $0.73$ & $0.35$ &
    $29.53$ & $0.84$ & $0.19$ \\
    QGAC~\cite{qgac} &
    $23.85$ & $0.64$ & $0.43$ &
    $28.01$ & $\mathbf{0.80}$ & $0.26$ &
    $31.28$ & $\mathbf{0.89}$ & $\mathbf{0.15}$ \\
    Ours (S) &
    $25.19$ & $0.70$ & $\mathbf{0.34}$ &
    $27.68$ & $0.78$ & $0.26$ &
    $30.98$ & $0.87$ & $\mathbf{0.15}$ \\
    Ours (A) &
    $\mathbf{25.90}$ & $\mathbf{0.72}$ & $\mathbf{0.34}$ &
    $\mathbf{28.37}$ & $\mathbf{0.80}$ & $\mathbf{0.25}$ &
    $\mathbf{31.58}$ & $0.88$ & $\mathbf{0.15}$ \\
    \bottomrule
  \end{tabular}
\end{table}

\section{JPEG Artifact Correction with DDRM}
For the case of no noise in the observation $\rvy$, the general DDRM process to sample from $p_\theta^{(t)}(\rvx_t | \rvx_{t+1}, \rvy)$ for linear inverse problems simplifies to be
\begin{align*}
    \rvx_{t}' & = f_\theta^{(t+1)}(\rvx_{t+1}) - \mH^{\dagger} \mH f_\theta^{(t+1)}(\rvx_{t+1}) + \mH^{\dagger} \rvy, \\
    \rvx_{t} & = \sqrt{\alpha_t} \left(\eta_b \rvx_{t}' + (1 - \eta_b) f_\theta^{(t+1)}(\rvx_{t+1}) \right) + \sqrt{1 - \alpha_t} \left(\eta \epsilon_t + (1 - \eta) \epsilon_\theta^{(t+1)}(\rvx_{t+1})\right)
\end{align*}
where $\mH^\dagger$ is the Moore-Penrose pseudo-inverse of $\mH$, $f_\theta^{(t+1)}(\rvx_{t+1})$ is the denoising model output at the previous step $t+1$, $\epsilon_\theta^{(t+1)}(\rvx_{t+1}) = \frac{\rvx_{t+1} - \sqrt{\alpha_{t+1}} f_\theta^{(t+1)}(\rvx_{t+1})}{\sqrt{1 - \alpha_{t+1}}}$ is the predicted noise value, $\eta$ and $ \eta_b$ are user-defined hyperparameters, and $\epsilon_t \sim \gN(0, \mI)$ is a standard Gaussian distributed vector. At a high level, we inject information about $\rvy$ via $\rvx'_t$ where we replace the values in the spectral domain with what we know from $\rvy$. The sampling process then aggregates $\rvx'_t$ (corrected from $\rvy$), $\rvx_{t+1}$ (the current input), and $f_\theta^{(t+1)}(\rvx_{t+1})$ (the denoiser output) to produce the value for the next iteration. 

While it seems that the above approach only works for linear $\mH$, its insights can actually be used for other non-linear inverse problems, such as JPEG artifact correction. We note that for a linear $\mH$, its pseudo-inverse $\mH^\dagger$ has two important properties:
\begin{enumerate}
    \item $\mH \mH^\dagger \mH = \mH$, \textit{i.e.}, taking the pseudo-inverse does not change the measurement.
    \item $\mH^\dagger \mH \rvx$ is ``close'' to $\rvx$, in the sense that $\mH^\dagger \mH \rvx$ provides a least squares solution for the problem $ \min_{\rvz} \norm{\rvz - \rvx}_2^2$ for all $\rvx$ when observing only $\rvy = \mH \rvx$ (but not $\rvx$).
\end{enumerate}
The above properties may exist for operators that are not linear. For example, if we treat $\mH$ as the JPEG encoding operator, then the JPEG decoding operator also satisfies these properties:
\begin{enumerate}
    \item The JPEG encoding introduces loss of information during the quantization and chroma subsampling stages. The remaining information is kept during JPEG decoding, and thus encoding it again will lead to the same result.
    \item The JPEG decoding method generally preserves visual similarity, so applying decoding after encoding should generate an image that is ``close'' to the original one.
\end{enumerate}
With this insight, we can simply perform JPEG restoration with DDRM with the update rule
\begin{align*}
    \rvx_{t}' & = f_\theta^{(t+1)}(\rvx_{t+1}) - \text{Decode}\left(\text{Encode}\left(f_\theta^{(t+1)}(\rvx_{t+1})\right)\right) + \text{Decode}\left(\rvy\right), \\
    \rvx_{t} & = \sqrt{\alpha_t} \left(\eta_b \rvx_{t}' + (1 - \eta_b) f_\theta^{(t+1)}(\rvx_{t+1})\right) + \sqrt{1 - \alpha_t} \left(\eta \epsilon_t + (1 - \eta) \epsilon_\theta^{(t+1)}(\rvx_{t+1})\right),
\end{align*}
which can be used in realistic settings as the quantization matrices are stored within the JPEG files.

\section{Experimental Results}

We evaluate our method on the ImageNet~\cite{imagenet} dataset, as it is diverse and represents real-world use cases.
Specifically, we evaluate on a $1000$ image subset of the ImageNet validation set named ImageNet-1K~\cite{dgp}. 
We utilize the diffusion model from~\cite{guided_diffusion}, trained on $256 \times 256$-pixel ImageNet training images, for a diffusion schedule of $1000$ timesteps.
In all of our experiments, we choose the hyperparameters $\eta = 1$, $\eta_b = 0.4$, and $20$ uniformly-spaced diffusion steps.
Additionally, since JPEG images generally preserve overall image contents, we find that we can perturb a JPEG-compressed image with noise and use it as an initialization for our sampling process at an intermediate step $t = 300$, similar to~\cite{meng2021sdedit}. This allows the sampling to provide more faithful reconstructions, avoiding the unnecessary randomness induced by starting at the initial timestep $T = 1000$.
However, as we use a probabilistic sampling scheme, randomness can still be expected in the results. In order to stabilize performance, we draw $8$ independent samples for each input and save the resulting average image.
We denote the first sample as \emph{``Ours (S)''} and the averaged image as \emph{``Ours (A)''}.

For our JPEG artifact correction experiments, we use the most common variant of JPEG~\cite{libjpeg}, which includes chroma subsampling and quantization matrices defined by a quality factor (QF) ranging from $1$ to $100$, $1$ being the most compressed, and $100$ being the most faithful to the original image.
Our method produces high quality reconstructions (see Figures \ref{fig:headline}, \ref{fig:extra_fig}). Moreover, when evaluated numerically on common metrics such as PSNR, SSIM~\cite{ssim}, and LPIPS~\cite{lpips}, our method provides a significant improvement over simple JPEG decoding, and its performance is favorable or comparable to a recent state-of-the-art JPEG artifact correction named QGAC~\cite{qgac}.
QGAC was specifically trained for JPEG restoration with QF $\in [10, 100]$, and as can be seen in \autoref{tab:imagenet}, it generalizes poorly for lower QF.
In contrast, our method generalizes well for all QF without JPEG-specific training.
We demonstrate its success by showing its compression rate-distortion curve in \autoref{fig:rd}.

Furthermore, our method is not limited to JPEG artifact correction, but can also be applied to similar non-linear inverse problems not covered by DDRM~\cite{ddrm}.
For instance, we consider the problem of image dequantization, where we attempt to recover high-quality reconstructions from images that were quantized below the standard $24$ bits per color.
As evident in \autoref{fig:quant}, our method generalizes well for image dequantization, owing to its problem-agnostic nature.

\section{Conclusion}
We propose a novel method for correcting JPEG compression artifacts using diffusion models. Our method extends DDRM~\cite{ddrm} beyond the linear case by generalizing the pseudo-inverse concept.
We perform evaluations on ImageNet-1K~\cite{imagenet,dgp} where our method performs on par with a state-of-the-art baseline in most cases, and exhibits generalization abilities for lower quality factors which specifically trained baselines do not possess.
Our method can further generalize beyond JPEG restoration, as we successfully demonstrate on the image dequantization problem.
It does so seamlessly, without retraining, and without problem-specific hyperparameter tuning.

\bibliography{main}
\bibliographystyle{plain}

\newpage
\appendix

\section{Additional Visual Results}
\begin{figure}[h]
    \centering
    \includegraphics[width=\textwidth]{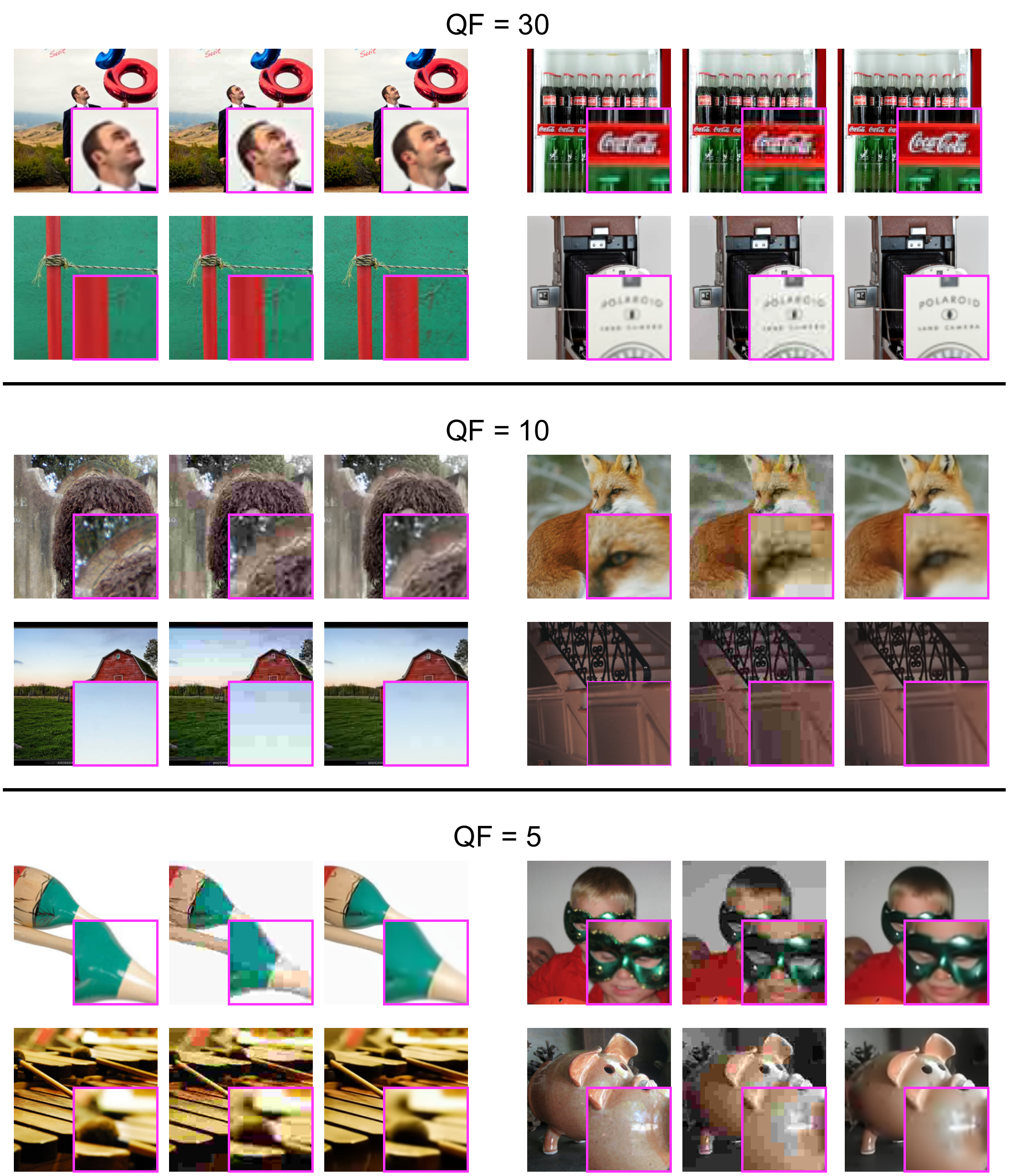}
    \caption{Triplets of original (ground-truth), JPEG compressed, and restored images. Across different quality factors (QF), our method successfully corrects artifacts of JPEG compression. Images are accompanied by a zoomed-in area in the bottom right corner to highlight specific artifact removals.}
    \label{fig:extra_fig}
\end{figure}

\end{document}